\documentclass[a4paper,11pt]{article}
\usepackage{enumitem}
\usepackage{pos}
\usepackage[utf8]{inputenc}
\usepackage[T1]{fontenc}
\usepackage{nicefrac}
\usepackage{amsmath}
\usepackage{braket}
\usepackage{graphicx}
\usepackage{tikz}
\usetikzlibrary{tikzmark}
\usepackage{multirow}
\usepackage{bm}
\usepackage{arydshln}

\definecolor{colorDG}{HTML}{008000}

\newcommand{\rn}{{\tt r}}
\newcommand{\Mn}{{\tt M}}
\newcommand{\R}{\text{Re}}
\newcommand{\I}{\text{Im}}

\title{Status of the Aligned Two Higgs Doublet Model in the low mass region}

\author*[a]{Anirban Karan}
\emailAdd{kanirban@ific.uv.es}

\author[a]{Antonio M. Coutinho}
\emailAdd{antonio.coutinho@ific.uv.es} 

\author[b]{V\'ictor Miralles,}
\emailAdd{victor.miralles@manchester.ac.uk}

\author[a]{Antonio Pich}
\emailAdd{antonio.pich@ific.uv.es}

\affiliation[a]{Instituto de F\'isica Corpuscular, CSIC -- Universitat de València, Parque Cient\'ifico, \\ Catedr\'atico Jos\'e Beltr\'an 2, E-46980 Paterna, Spain}

\affiliation[b]{Department of Physics and Astronomy, University of Manchester, Oxford Road, Manchester M13 9PL, United
Kingdom}

\abstract{The Two Higgs Doublet Model (2HDM) is a simple extension of the Standard Model (SM),
which provides a rich and very interesting phenomenology.
To remove the undesirable flavour-changing neutral currents (FCNCs), 
an additional $\mathcal Z_2$ symmetry is usually imposed into the 2HDM. However, FCNCs can be avoided in a much more general way by assuming a similar Yukawa structure for the two scalar doublets.
The model with this intriguing feature is termed the Aligned Two Higgs Doublet Model (A2HDM). The phenomenological constraints on the A2HDM are much weaker than the ones on the usual $\mathcal Z_2$ models, opening a broader range of possible scenarios. Moreover,
the A2HDM also provides a generic framework to study, as particular cases, the different varieties of $\mathcal Z_2$-symmetric 2HDMs. Here, we illustrate a global fit of the A2HDM using the package HEPfit. We study the possibility of having new scalar particles lighter than the SM Higgs. 
For this global fit we perform a Bayesian analysis, including stability and perturbativity bounds, flavour and electroweak precision observables, and scalar (and pseudoscalar) searches at LEP and LHC.}

\FullConference{42nd International Conference on High Energy Physics (ICHEP2024)\\
18-24 July 2024\\
Prague, Czech Republic\\}

\begin{document}
\maketitle

\section{The Aligned Two Higgs Doublet Model}
The 2HDM \cite{Branco:2011iw} extends the SM with one additional scalar doublet having hypercharge $1/2$. 
One can always work in the so-called ``Higgs basis'' where only one of the scalar doublets acquires the vev of $v=246$ GeV. Apart from the Goldstone bosons $(G^0,G^\pm)$, that give masses to the $Z$ and $W^\pm$ bosons, this model contains one charged ($H^+$) and three neutral ($S_{\{1,2,3\}}$) scalars, providing a very rich phenomenology. In the CP-conserving case, the CP-even scalars $S_{1,2}$ mix together to produce the physical states $(h,H)$ while the CP-odd state $S_3$ remains unmixed:
{\small
\begin{equation}
\Phi_1=\frac{1}{\sqrt 2}\begin{pmatrix}
\sqrt 2\;G^+\\
S_1+v+i\, G^0
\end{pmatrix}\, ,\;\;\Phi_2=\frac{1}{\sqrt 2}\begin{pmatrix}
\sqrt 2\;H^+\\
S_2+i\, S_3
\end{pmatrix}
\;\;\longrightarrow\;\;
\begin{pmatrix}
h\\H
\end{pmatrix}=\begin{pmatrix}
\cos\tilde{\alpha}& \sin\tilde{\alpha}\\ -\sin\tilde{\alpha}&\cos\tilde{\alpha}
\end{pmatrix}\begin{pmatrix}
S_1\\S_2
\end{pmatrix}\;\;\text{and}\;\; A=S_3\, .
\end{equation}
}
The scalar potential in the Higgs basis can generically be expressed as:
\begin{align}
\label{eq:pot}
\mathcal V&=\mu_1\,\Phi_1^\dagger \Phi_1+\mu_2\,\Phi_2^\dagger \Phi_2+ \Big[\mu_3\,\Phi_1^\dagger \Phi_2+h.c.\Big]+\frac{\lambda_1}{2}\,(\Phi_1^\dagger \Phi_1)^2+\frac{\lambda_2}{2}\,(\Phi_2^\dagger \Phi_2)^2+\lambda_3\,(\Phi_1^\dagger \Phi_1)(\Phi_2^\dagger \Phi_2)\nonumber\\
&+\lambda_4\,(\Phi_1^\dagger \Phi_2)(\Phi_2^\dagger \Phi_1)+\Big[\Big(\frac{\lambda_5}{2}\,\Phi_1^\dagger \Phi_2+\lambda_6 \,\Phi_1^\dagger \Phi_1 +\lambda_7 \,\Phi_2^\dagger \Phi_2\Big)(\Phi_1^\dagger \Phi_2)+ \mathrm{h.c.}\Big]\, ,
\end{align}
where all couplings are assumed real in the CP-conserving case. 
The minimization condition $v^2=-2\mu_1/\lambda_1=-2\mu_3/\lambda_6$ reduces
the number of independent parameters in the scalar sector to nine: $\mu_2\, , v\, , \lambda_{\{1,..,7\}}$. Furthermore, with the help of the following relations,
\begin{align}
\label{eq:mix_ang_lam}
&\tan\tilde\alpha=\frac{M_h^2 -v^2\,\lambda_1}{v^2\,\lambda_6}=\frac{v^2\,\lambda_6}{v^2\,\lambda_1-M_H^2}\, , \quad M_{h,H}^2=\frac{1}{2}\,(\Sigma\mp\Delta),\quad M_A^2=M_{H^\pm}^2+\frac{v^2}{2}\,(\lambda_4-\lambda_5)\, , \nonumber\\
&M_{H^\pm}^2=\mu_2+\frac{\lambda_3}{2}\,v^2 ,  \text{ with } \;\Sigma=M_{H^\pm}^2+\Big(\lambda_1+\frac{\lambda_4}{2}+\frac{\lambda_5}{2}\Big)\,v^2\; \text{ and }\; \Delta=\sqrt{\big(\Sigma-2\lambda_1 v^2\big)^2+4\,\lambda_6^2\,v^4}\, ,
\end{align}
a few parameters in the above list can be replaced by the physical masses and the mixing angle. Since $M_h$ and $v$ are known quantities, we are left with seven independent parameters in the scalar sector: $M_{\{H^\pm,H,A\}},\,\lambda_{\{2,3,7\}}$ and $\tilde{\alpha}$. On the other hand, from the kinetic term,
the couplings of the neutral scalars with the gauge bosons can be written as $
g_{hVV}=\cos\tilde{\alpha}\;g_{hVV}^{SM}\, , \; g_{HVV}=-\sin\tilde{\alpha}\; g_{hVV}^{SM}\;\;\text{and} \;\; g_{AVV}=0$ \
(with $VV\equiv W^+W^-, ZZ$).

In the basis of fermion mass eigenstates, the 
Yukawa Lagrangian takes the form \big($\tilde\Phi_a\equiv i \tau_2 \Phi^*_a$\big):  
\begin{equation}
    -\mathcal L_Y=\Big(\frac{\sqrt 2}{v}\Big)\;\Big\{\bar Q_L (M_u \tilde\Phi_1 + Y_u \tilde\Phi_2)u_R + \bar Q_L (M_d \Phi_1 + Y_d \Phi_2)d_R + \bar L_L (M_\ell \Phi_1 + Y_\ell \Phi_2)\ell_R + \mathrm{h.c.} \Big \} \, ,
\end{equation}
where $Y_f$ are arbitrary $3\times3$ matrices 
and $M_f$ $(f\equiv u,d,\ell)$ the diagonal mass matrices of the fermions. 
Tree-level FCNCs can be avoided by imposing $Y_f=\varsigma_f \, M_f$ (with real $\varsigma_f$ in the CP-conserving case) \cite{Pich:2009sp}; the interaction part of the Yukawa Lagrangian becomes then:
{\small\begin{equation}
-\mathcal L_Y\supset\sum\Big({y_f^{\varphi^0_i}}\big/{v}\Big)\,\varphi^0_i\,\Big[\bar f M_f \mathcal{P}_R f\Big]+\big(\sqrt 2\big/{v}\big)\, H^+\,\Big[\bar u\,\big\{\varsigma_d V M_d \mathcal{P}_R-\varsigma_u M_u^\dagger V\mathcal{P}_L\big\}\, d+\varsigma_\ell\, \bar \nu M_\ell \mathcal P_R \ell\Big] + \mathrm{h.c.}\, ,
\end{equation}}
where $\mathcal P_{L,R}=(1\mp\gamma^5)/2$\,, $\varphi_i^0\,\equiv \{h,H,A\}$, $V$ is the CKM matrix and the Yukawa couplings are:
\begin{align}
&y_{f}^H=-\sin\tilde\alpha+\varsigma_{f}\,\cos \tilde\alpha\, , \qquad y_{f}^h=\cos\tilde\alpha+\varsigma_{f}\,\sin \tilde\alpha\, , \qquad
y_{u}^A= - i\varsigma_{u}\, , \qquad y_{d,\ell}^A=  i\varsigma_{d,\ell}\, .
\label{eq:Higgs_yuk_up}
\end{align}
 One can retrieve the usual $\mathcal{Z}_2$-symmetric 2HDM cases from this A2HDM scenario by simply imposing $\mu_3=\lambda_6=\lambda_7=0$, together with the following conditions: 
\begin{align}
&\text{Type I:\;\;} \varsigma_{u}=\varsigma_d=\varsigma_\ell=\cot\beta,\quad \text{Type II:\;\;} \varsigma_{u}=-\varsigma_d^{-1}=-\varsigma_\ell^{-1}=\cot\beta\, ,\quad 
\text{Inert:\;\;}\varsigma_{u}=\varsigma_d=\varsigma_\ell=0\, ,
\nonumber\\
&\text{Type X:\;\;} \varsigma_{u}=\varsigma_d=-\varsigma_\ell^{-1}=\cot\beta 
\qquad \text{and}\qquad
\text{Type Y:\;\;} \varsigma_{u}=-\varsigma_d^{-1}=\varsigma_\ell=\cot\beta \, .
\end{align}
Though radiative corrections introduce some misalignment of the Yukawa couplings at higher loops, this effect can be shown to be well-below the current experimental reach \cite{Braeuninger:2010td,Jung:2010ik,Penuelas:2017ikk}.

\section{Constraints imposed on the model}
\begin{enumerate}[label=(\alph*)]
\setlength\itemsep{0mm}
    \item \textbf{Vacuum Stability:} The scalar potential can be written as
   $\mathcal V=-\,\Mn_\mu\,{\rn}^\mu + \nicefrac{1}{2}\,\Lambda^{\mu}_{\phantom{\mu}\nu}\, \rn_\mu\,\rn^\nu$, with 
$\rn^\mu\, =\, \big\{|\Phi_1|^2+|\Phi_2|^2, \, 2\,\R (\Phi_1^\dagger\Phi_2), \, 2\,\I (\Phi_1^\dagger\Phi_2), \, |\Phi_1|^2-|\Phi_2|^2\big\}$  a Minkowskian 4-vector.
The ``bounded from below'' condition 
is ensured if: 1) all the eigenvalues of $\Lambda^{\mu}_{\phantom{\mu}\nu}$ are real, and 2) the ``timelike'' eigenvalue $\Lambda_0$  is larger than the three ``spacelike'' eigenvalues $\Lambda_{1,2,3}$, along with $\Lambda_0>0$ \cite{Ivanov:2015nea}. Moreover, the vacuum can be guaranteed to be a stable neutral minimum by imposing: i) $D>0$, or ii) $D<0$ with 
$\xi =\big(M_{H^\pm}^2\big/v^2\big) >\Lambda_0$, where $D=\mathrm{Det}[\xi\, {\mathbb I_4}-\Lambda^{\mu}_{\phantom{\mu}\nu}]$ 
\cite{Ivanov:2015nea}. The masses and quartic couplings of the scalars become restricted by these conditions.

    \item \textbf{Perturbativity:}
To guarantee perturbative unitarity, one first constructs
the matrix of tree-level partial-wave amplitudes 
for all the $2\to2$ scatterings involving scalars and Goldstones 
and then demands the eigenvalues of the S-wave amplitudes at very high energy to satisfy $(a_0^{0})^2\leq \frac{1}{4}$, where
$(\mathbf{a_0})_{i,f}=\frac{1}{16\pi s}\int_{-s}^{0} dt \;\mathcal M_{i\to f}(s,t)$ \cite{Ginzburg:2005dt}. These constraints also restrict the masses and quartic couplings of the scalars to a great extent. On the other hand, perturbativity in the fermionic sector is ensured by demanding the couplings of the charged Higgs to pairs of fermions to be less than unity, which implies: $|\varsigma_f|<v/\big(\sqrt 2 m_f\big)$.

    \item \textbf{Oblique Parameters:} S, T and U
    are very sensitive to the existence of additional scalars beyond the SM (BSM) \cite{Celis:2013rcs}. The values of these oblique parameters as well as the correlations among them are extracted from a fit to the electroweak precision observables,  removing $R_b\equiv \Gamma(Z\rightarrow b\bar{b})/\Gamma(Z\rightarrow \rm{hadrons})$ since it gets also modified by the new scalars. The oblique parameters restrict the mass splitting between the charged and neutral BSM particles.


    \item \textbf{Flavour observables:} The BSM scalars generate significant contributions to loop-induced observables like $\Delta M_{B_s}$, $\mathcal{B}(B\to X_s\gamma)$, $\mathcal{B}(B_s\to \mu\mu)$ and $R_b$ \cite{Jung:2010ik}. By the same token, various observables involving tree-level processes such as $\mathcal{B}(B\to \tau \nu)$, $\mathcal{B}(D_{(s)}\to \tau\nu)$, $\mathcal{B}(D_{(s)}\to \mu\nu)$, $\frac{\Gamma(K\to \mu\nu)}{\Gamma(\pi\to \mu\nu)}$ and $\frac{\Gamma(\tau\to K\nu)}{\Gamma(\tau\to \pi\nu)}$ also gain contributions from these extra scalars \cite{Jung:2010ik}. The experimentally measured values of these observables \cite{ParticleDataGroup:2024cfk} restrict the parameters $\tilde\alpha$ and $ \varsigma_f$ and the masses of the scalars (specifically that of the charged scalar).

    \item \textbf{Higgs signal strengths:} Different production and decay channels of the SM Higgs have been measured with good precision at the LHC. In this model, the couplings of the 125 GeV Higgs ($h$) with electroweak gauge bosons and fermions alter as $g_{hVV}=\cos\tilde{\alpha}\;g_{hVV}^{SM}$ and $C_{hff}=y_f^h\;g_{hff}^{SM}$ \cite{Celis:2013rcs}. As such, the experimentally measured Higgs signal strengths \cite{ParticleDataGroup:2024cfk}, 
    $\mu_{XY} = \sigma(pp\to h)\;\mathcal{B} (h\to XY)/ [\sigma(pp\to h)\;\mathcal{B} (h\to XY)]_{SM}$,
    constrain the alignment parameters $\varsigma_f$ and the mixing angle $\tilde \alpha$.

    \item \textbf{Collider searches:} Experiments at LEP \cite{ALEPH:2006tnd,ALEPH:2013htx} and LHC \cite{CMS,ATLAS} have performed extensive searches for BSM scalars. Additionally, if these particles are lighter than the SM Higgs, they might contribute to the widths (or invisible widths) of $W, Z$ and $h$, which are very well measured \cite{ParticleDataGroup:2024cfk}. Thus, the masses and couplings of the extra scalars receive restrictions from all these collider searches \cite{Celis:2013ixa}.
\end{enumerate}
    
\begin{figure}[t!]
    \centering
   \scalebox{1.1}{
   \hspace{-11mm}
    \begin{tabular}{|c||c||c|}
    \hline
     \includegraphics[scale=0.25]{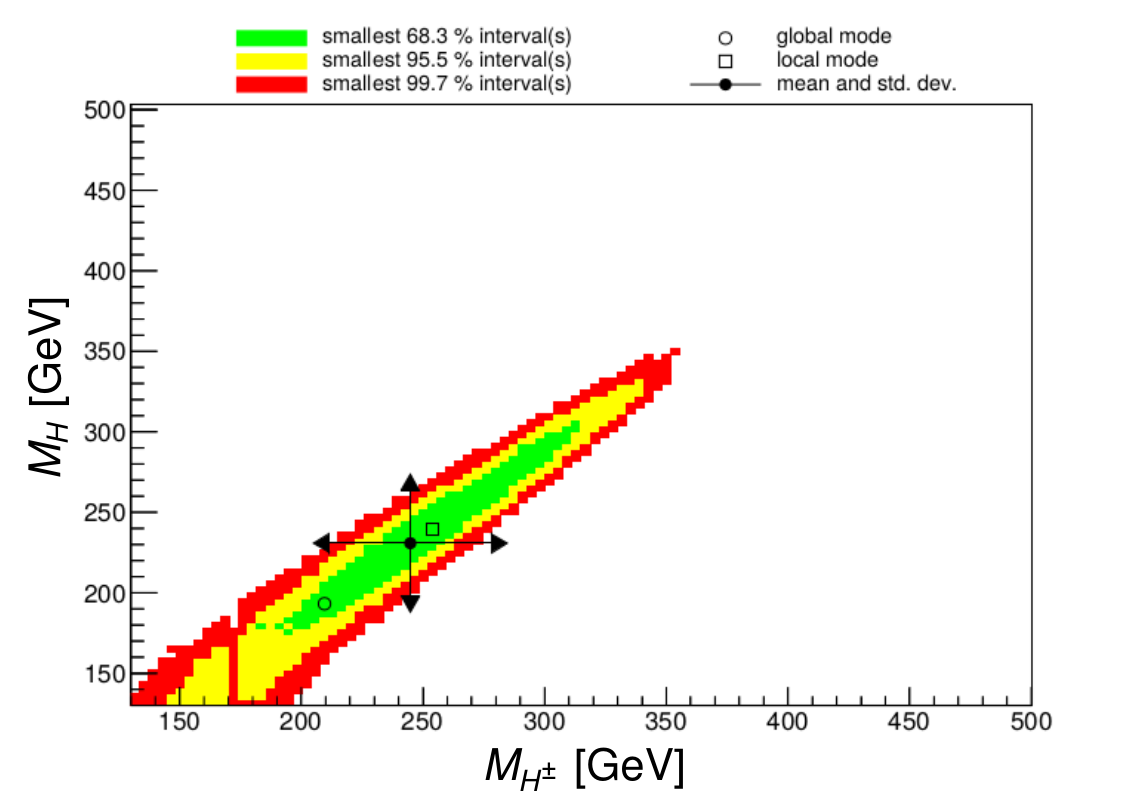}&
\includegraphics[scale=0.25]{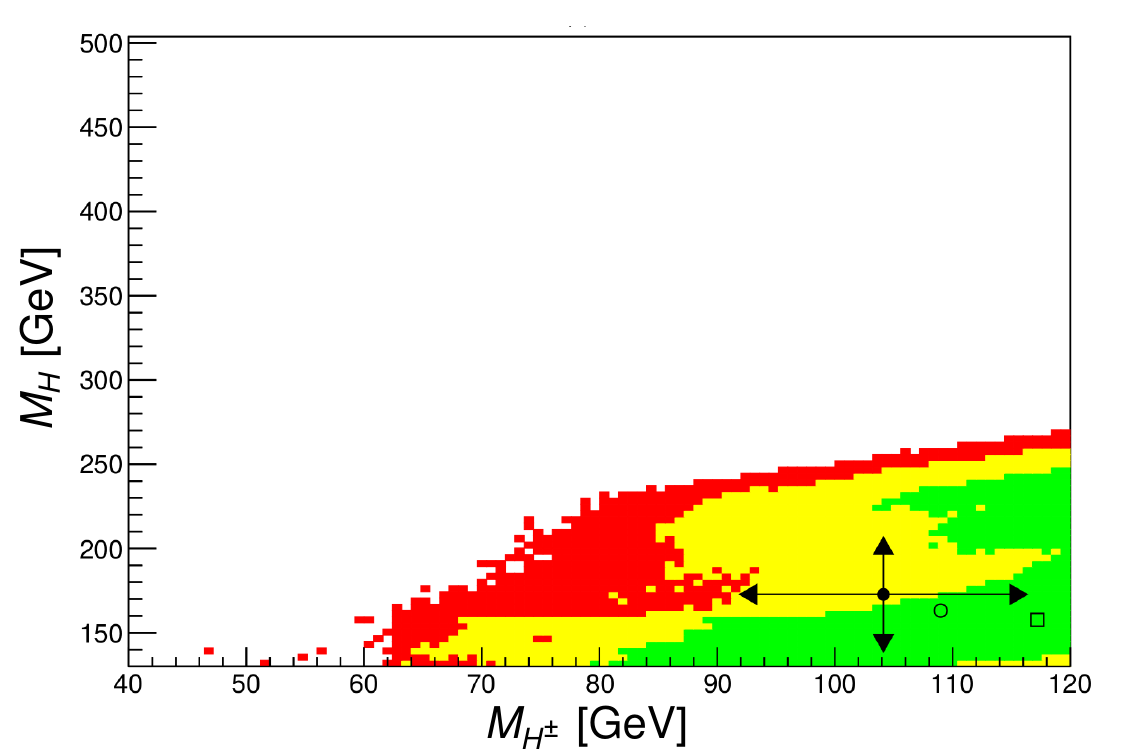}&
\includegraphics[scale=0.25]{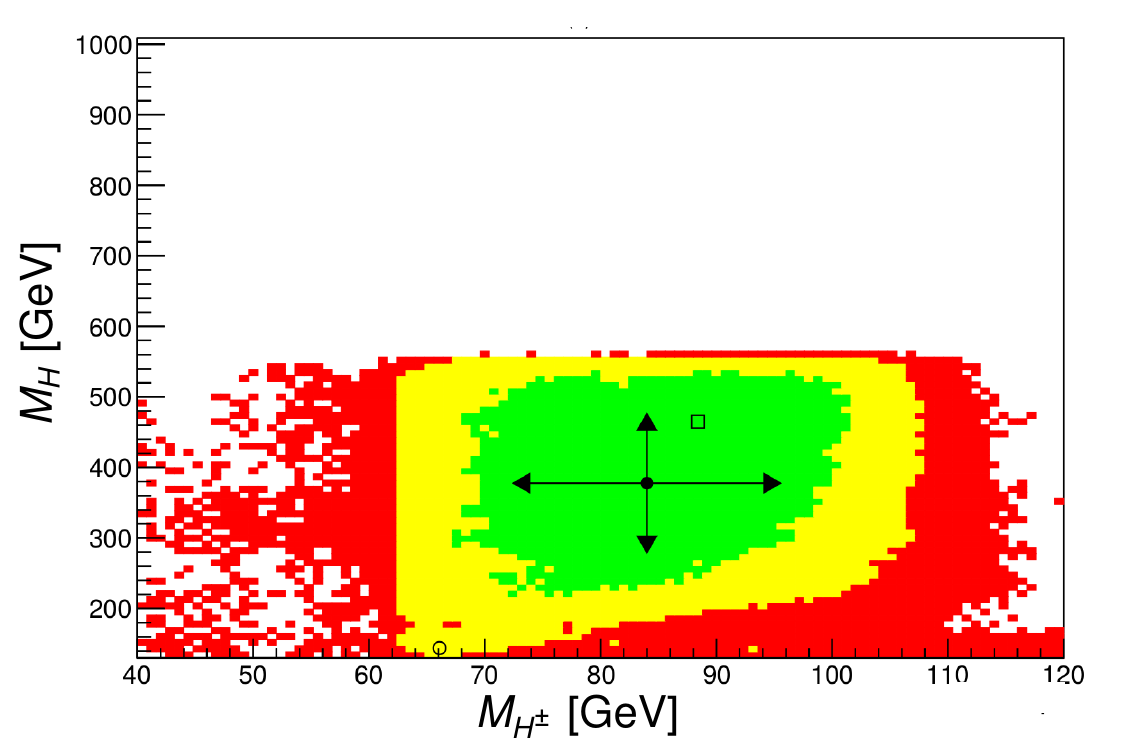}\\
\hdashline
\includegraphics[scale=0.25]{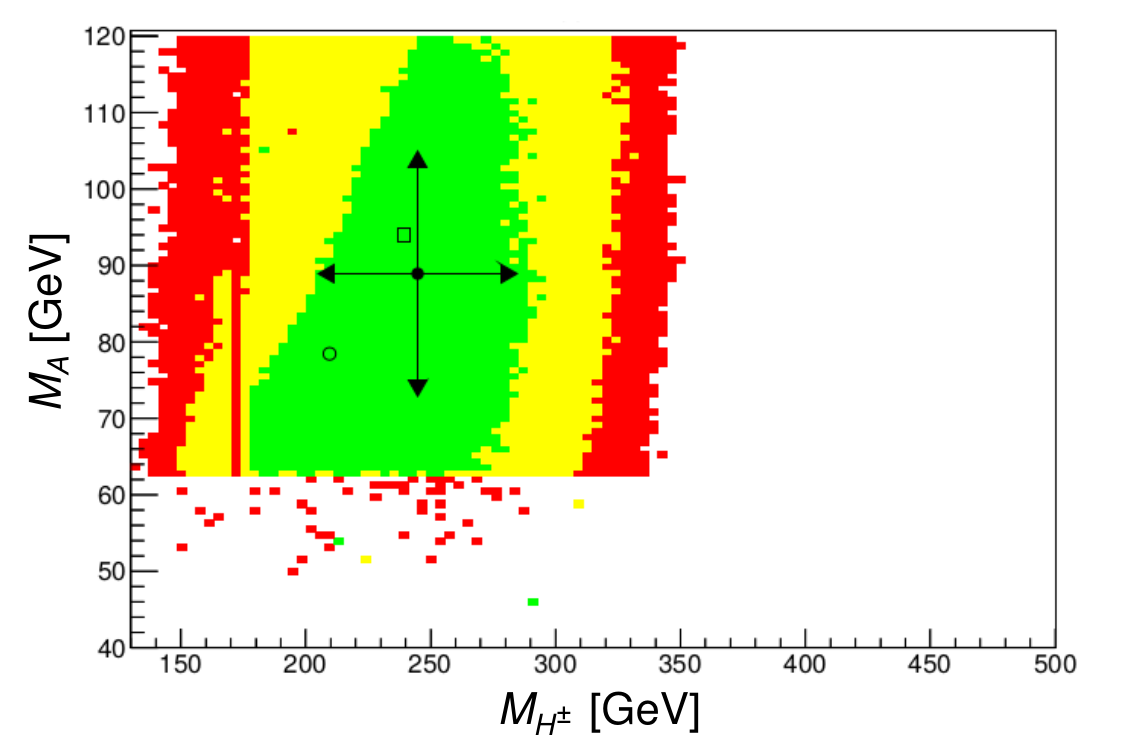}&
	\includegraphics[scale=0.25]{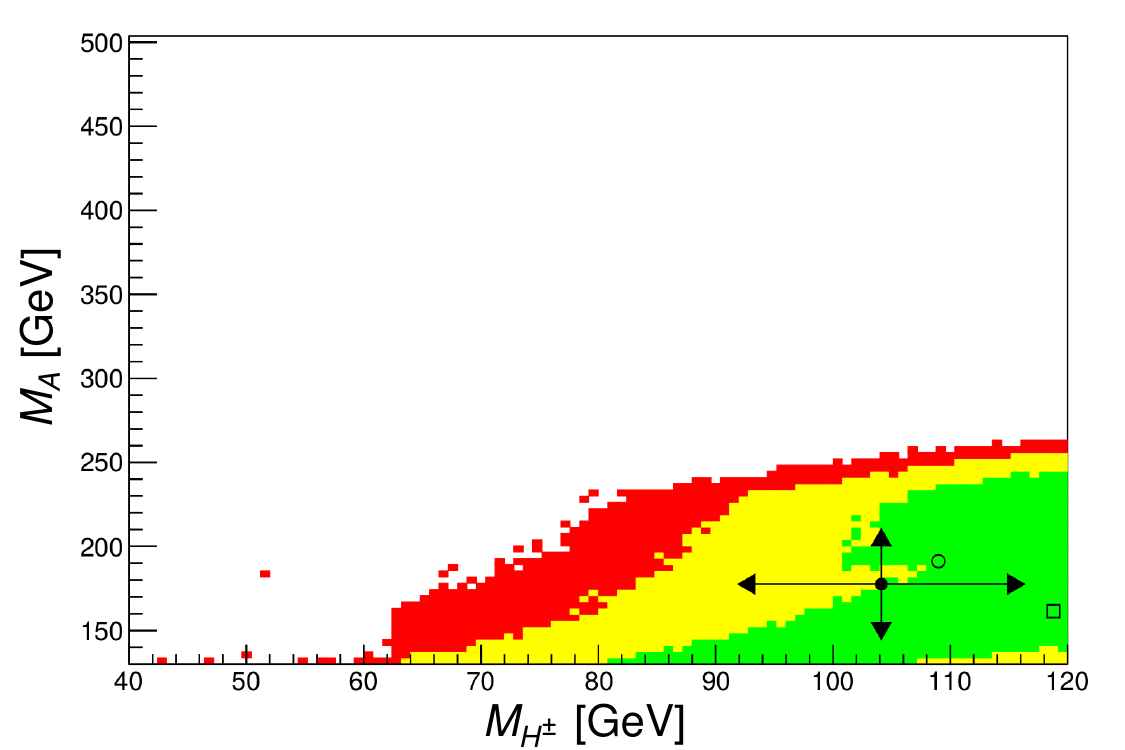}&
	\includegraphics[scale=0.25]{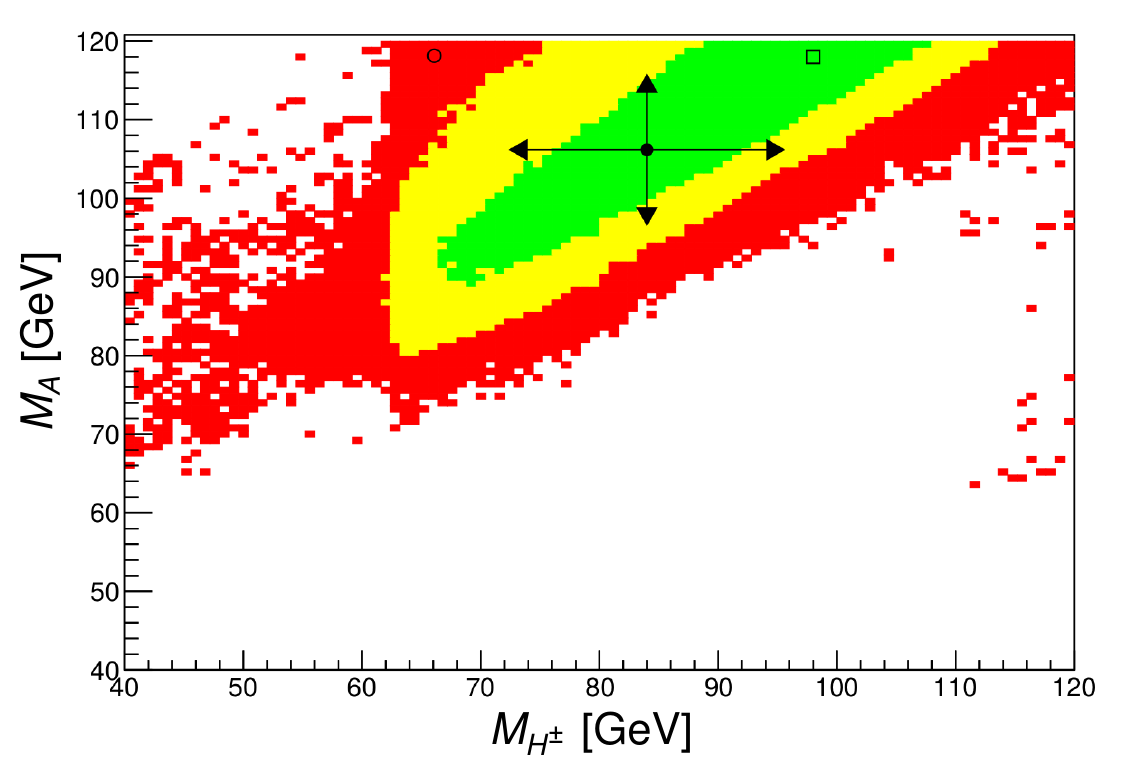}\\
\hdashline
\includegraphics[scale=0.25]{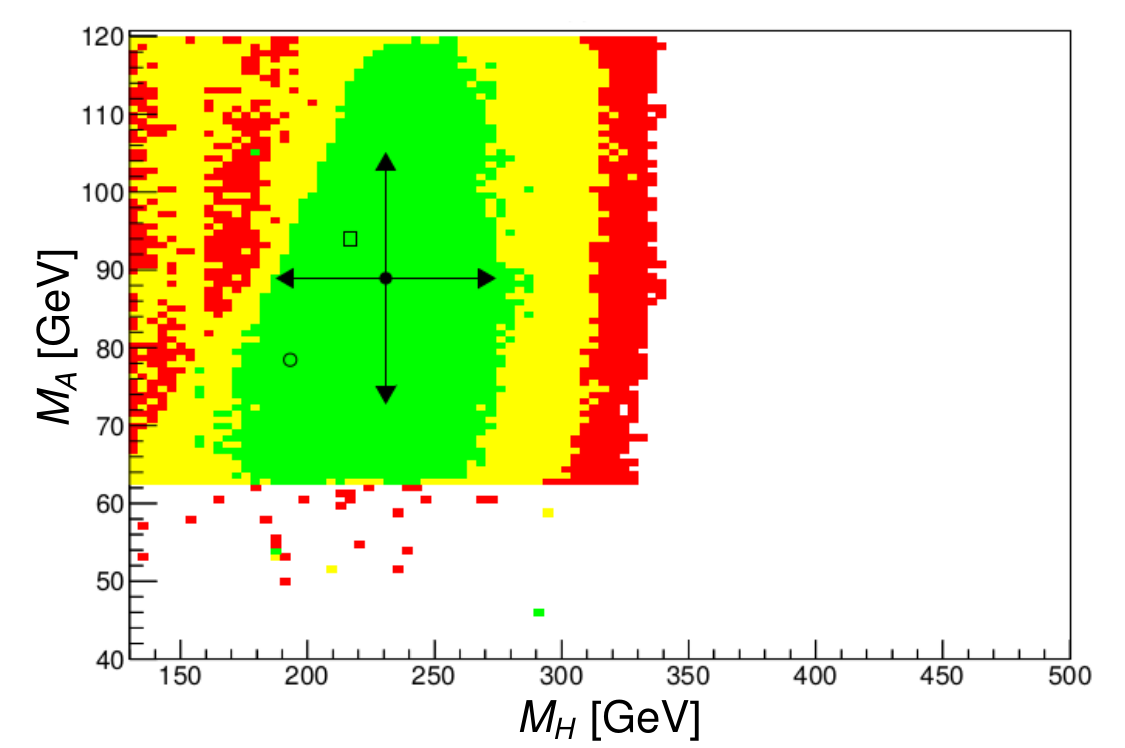}&
	\includegraphics[scale=0.25]{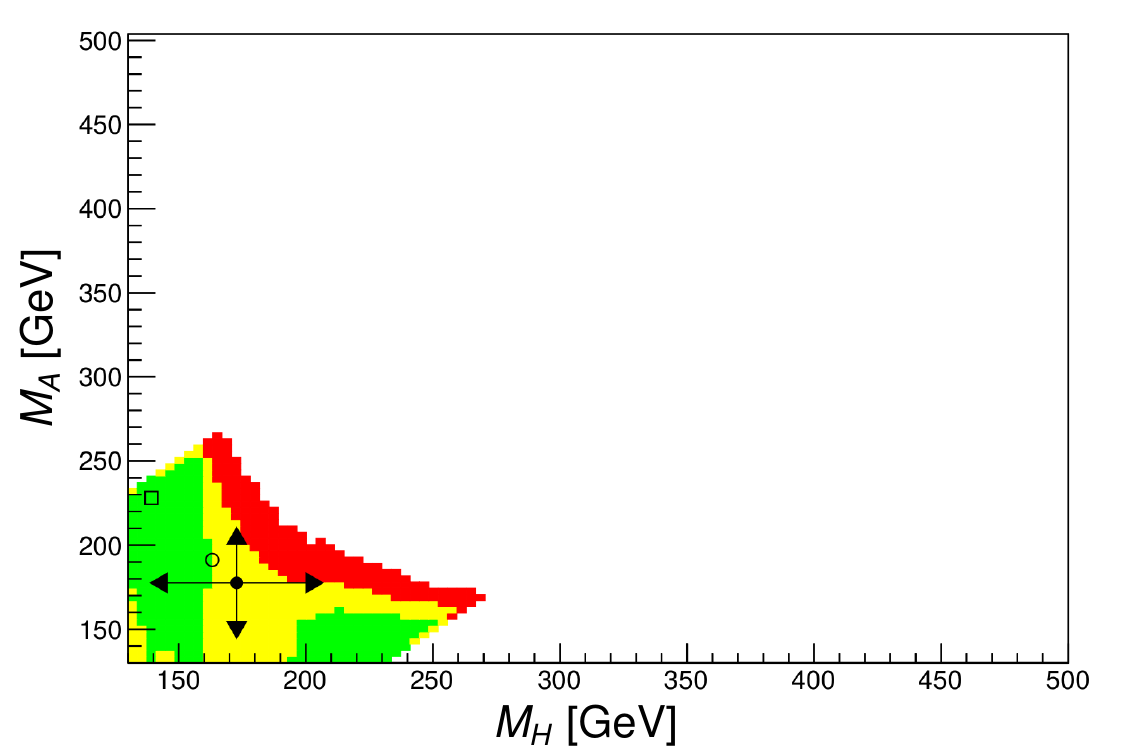}&
	\includegraphics[scale=0.25]{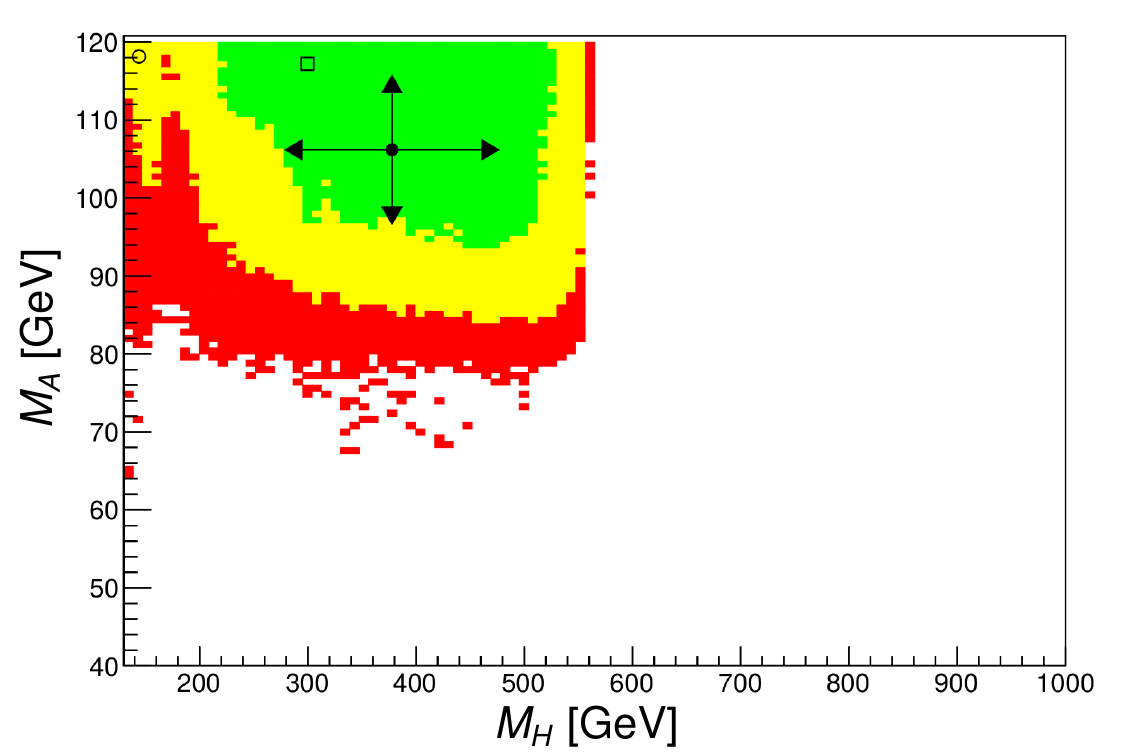}\\
\hline
Light $A$ & Light $H^\pm$ & Light $A$ \& $H^\pm$\\
\hline
    \end{tabular}
 }   
    \caption{Correlations among the masses of the additional scalars in the three cases: a) light pseudoscalar, b) light charged scalar, c) pseudoscalar and charged scalar both lighter than the SM Higgs. The red, yellow and green regions represent $1\sigma$, $2\sigma$ and $3\sigma$ regions, respectively (LEP results are not included).}
    \label{fig:mass}
\end{figure}

\section{Global fits}
\begin{figure}
    \centering
   \scalebox{1.1}{
   \hspace{-11mm}
    \begin{tabular}{|c||c||c|}
    \hline
     \includegraphics[scale=0.25]{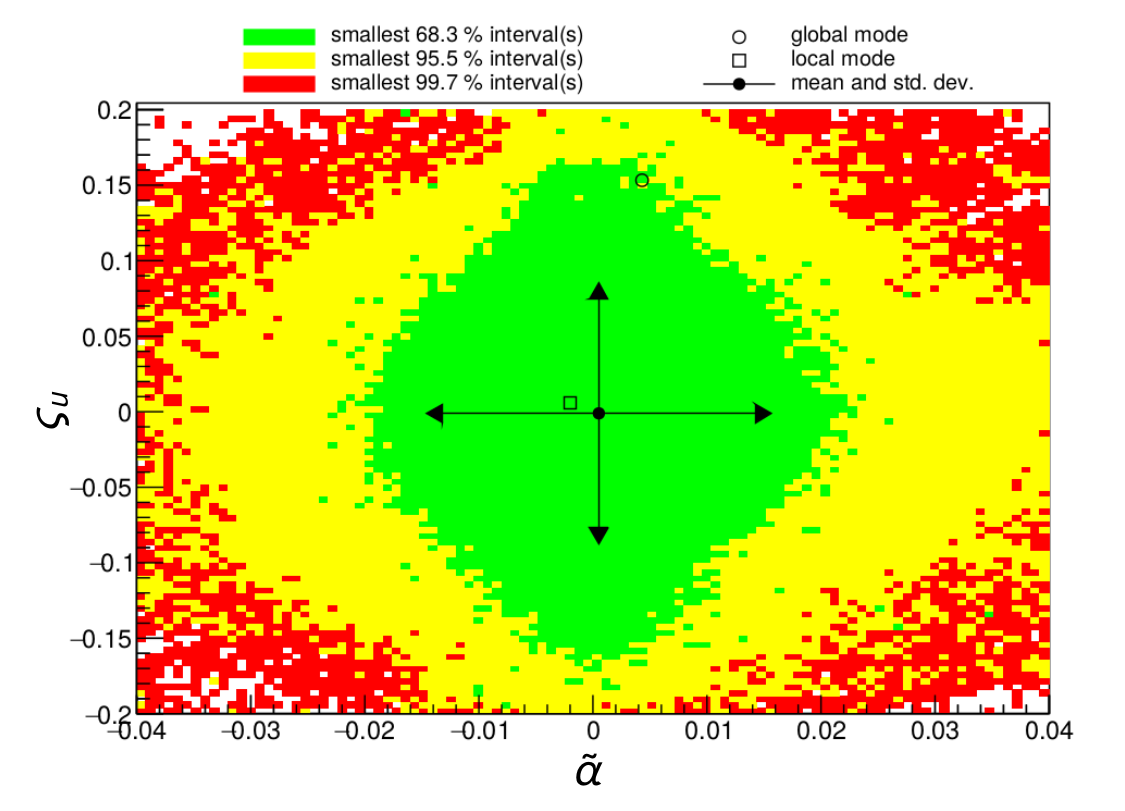}&
\includegraphics[scale=0.25]{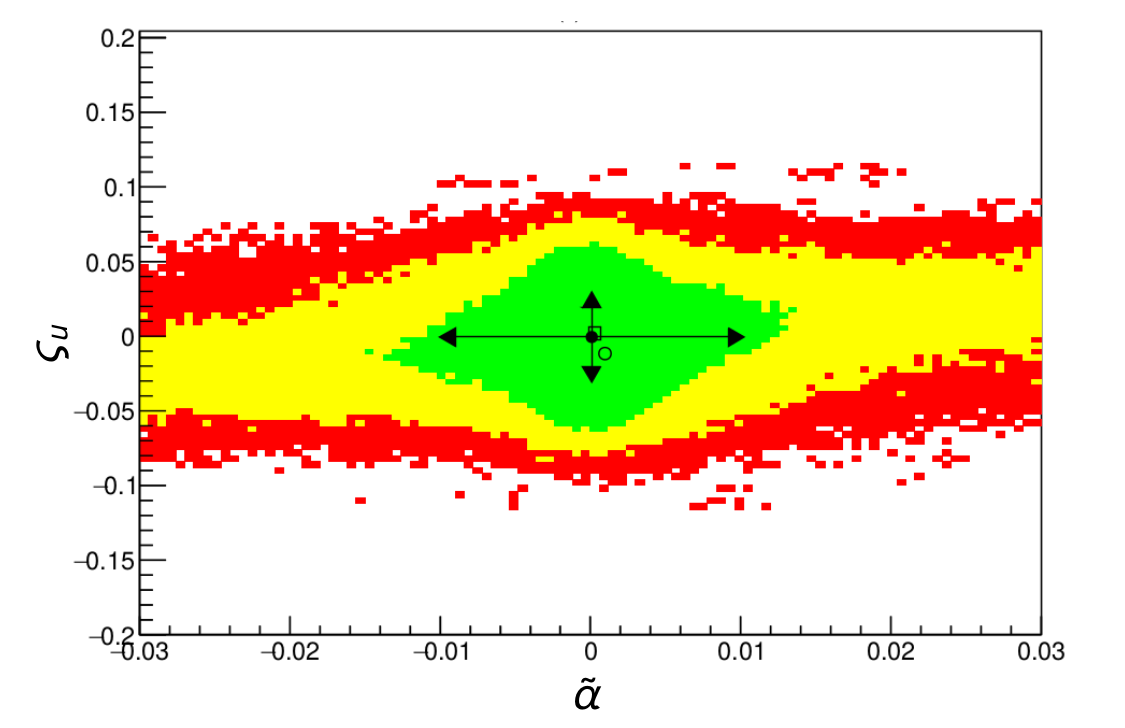}&
\includegraphics[scale=0.25]{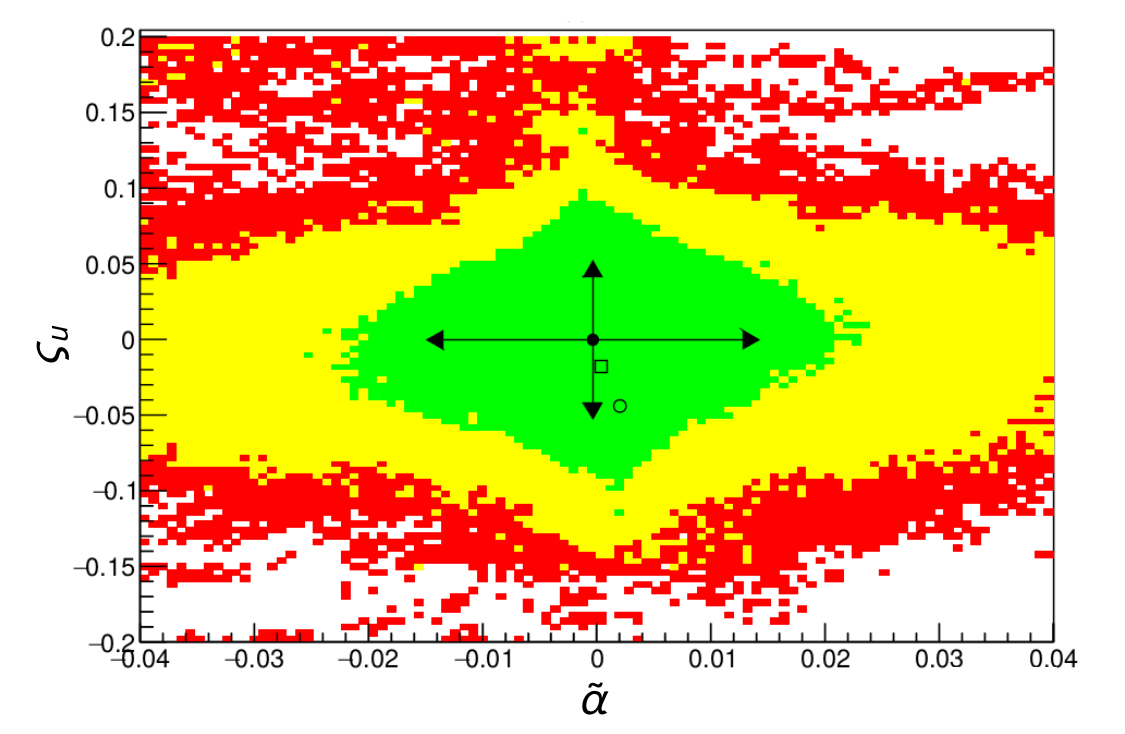}\\
\hdashline
\includegraphics[scale=0.25]{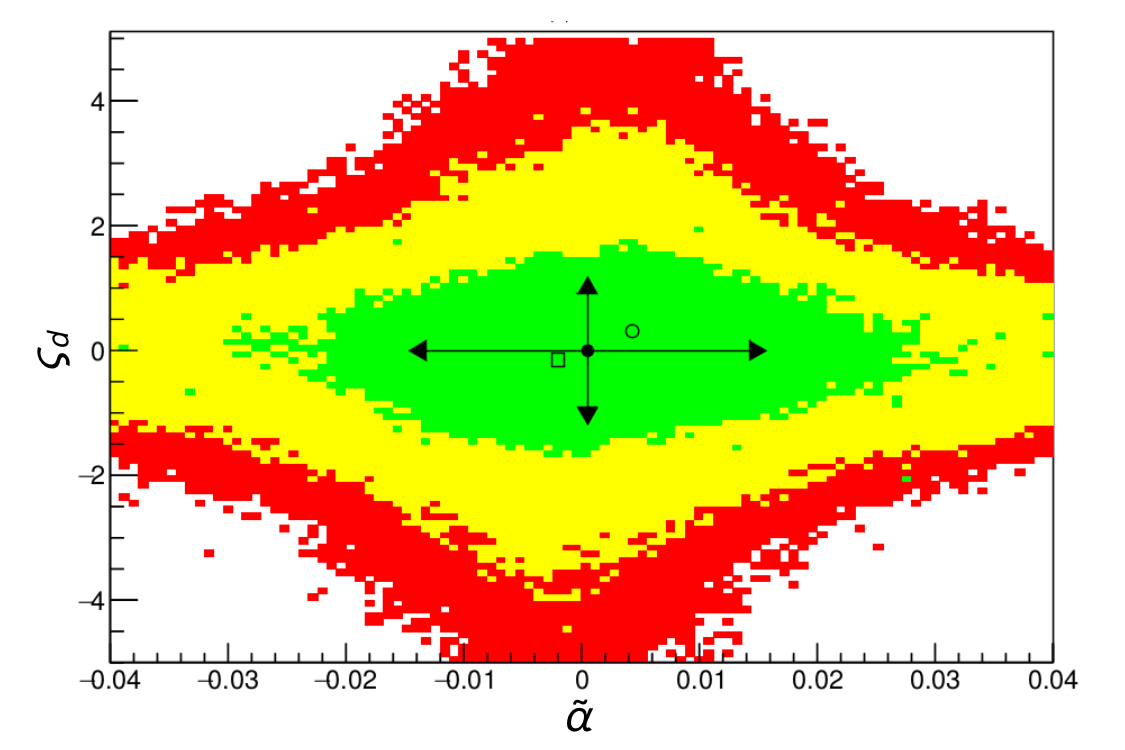}&
	\includegraphics[scale=0.25]{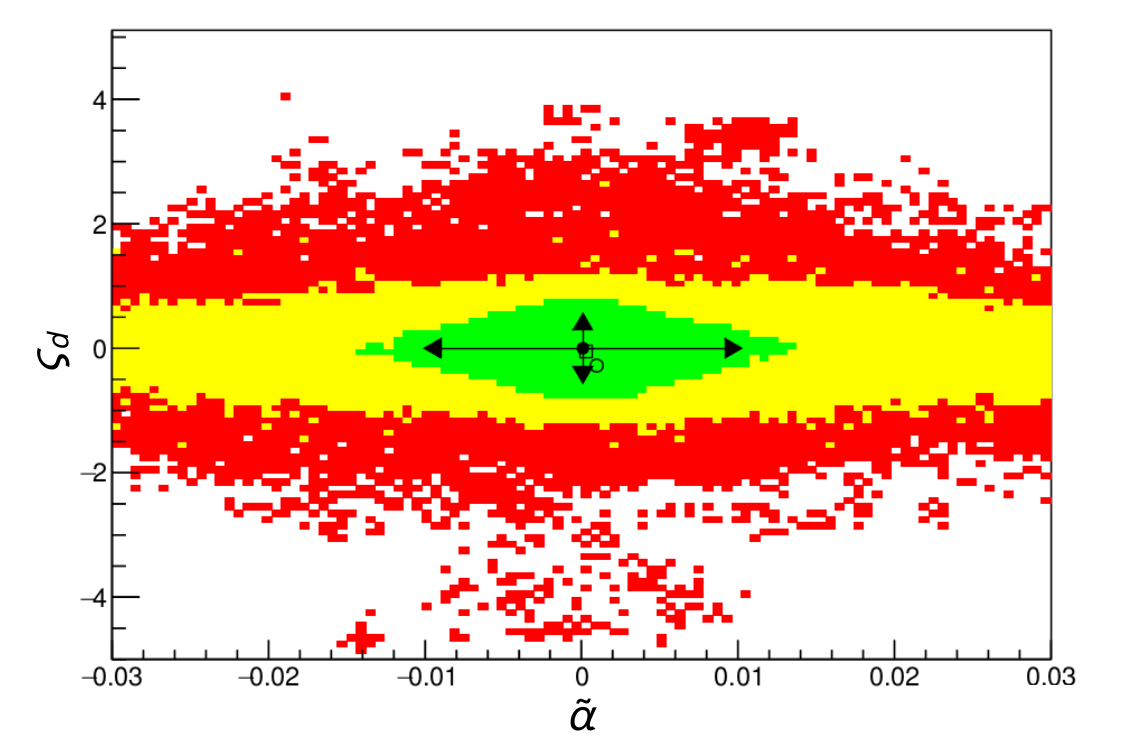}&
	\includegraphics[scale=0.25]{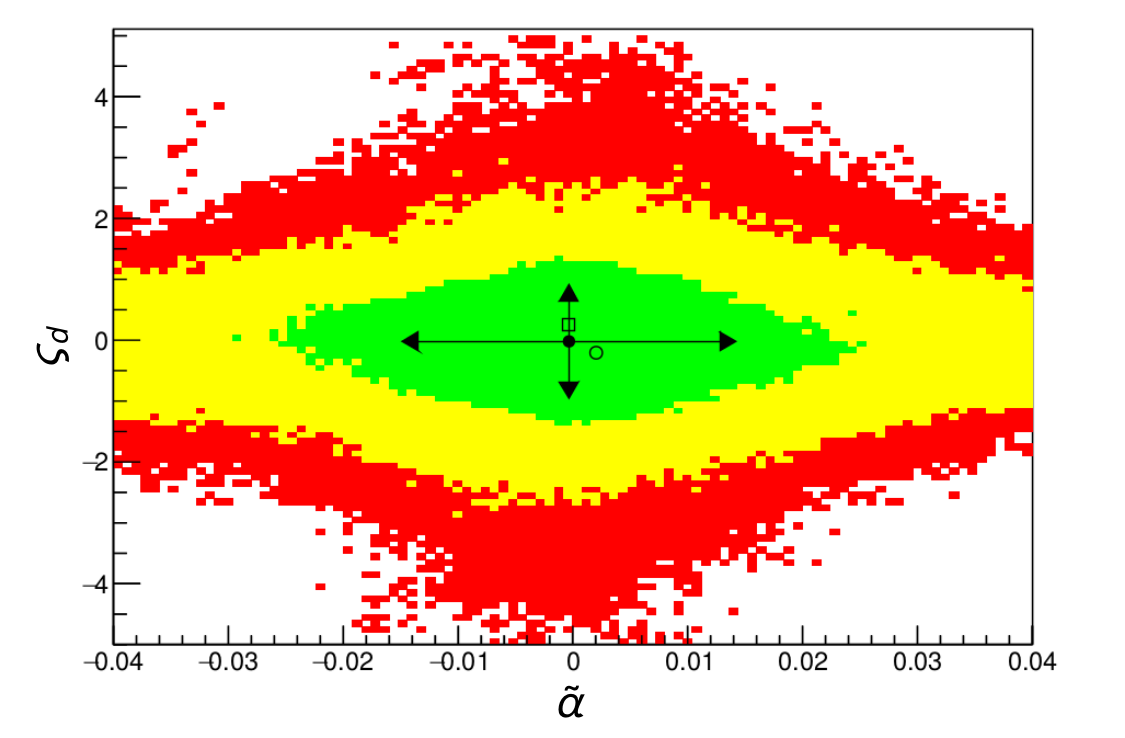}\\
\hdashline
\includegraphics[scale=0.25]{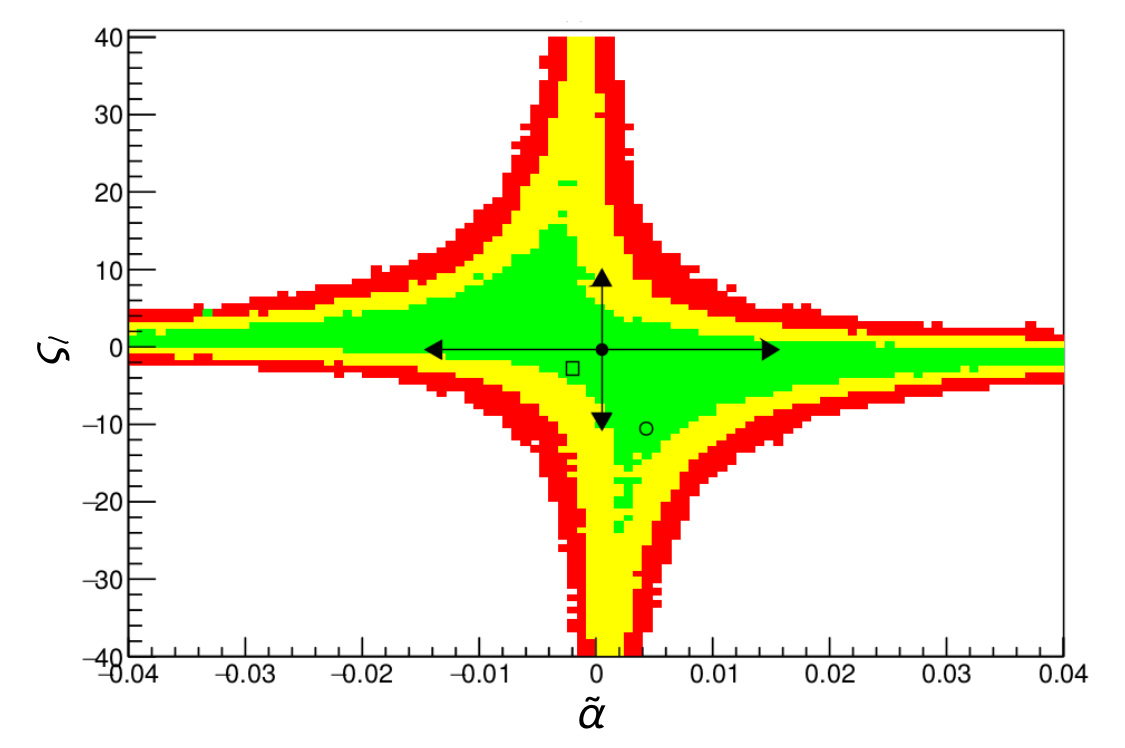}&
	\includegraphics[scale=0.25]{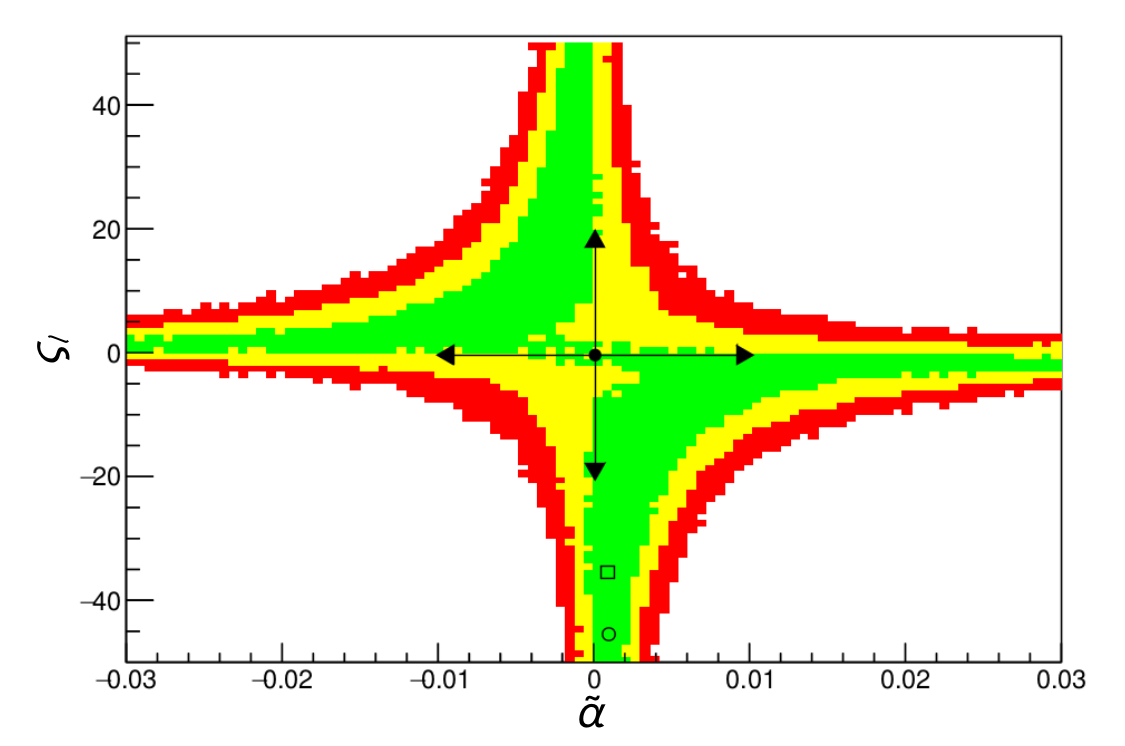}&
	\includegraphics[scale=0.25]{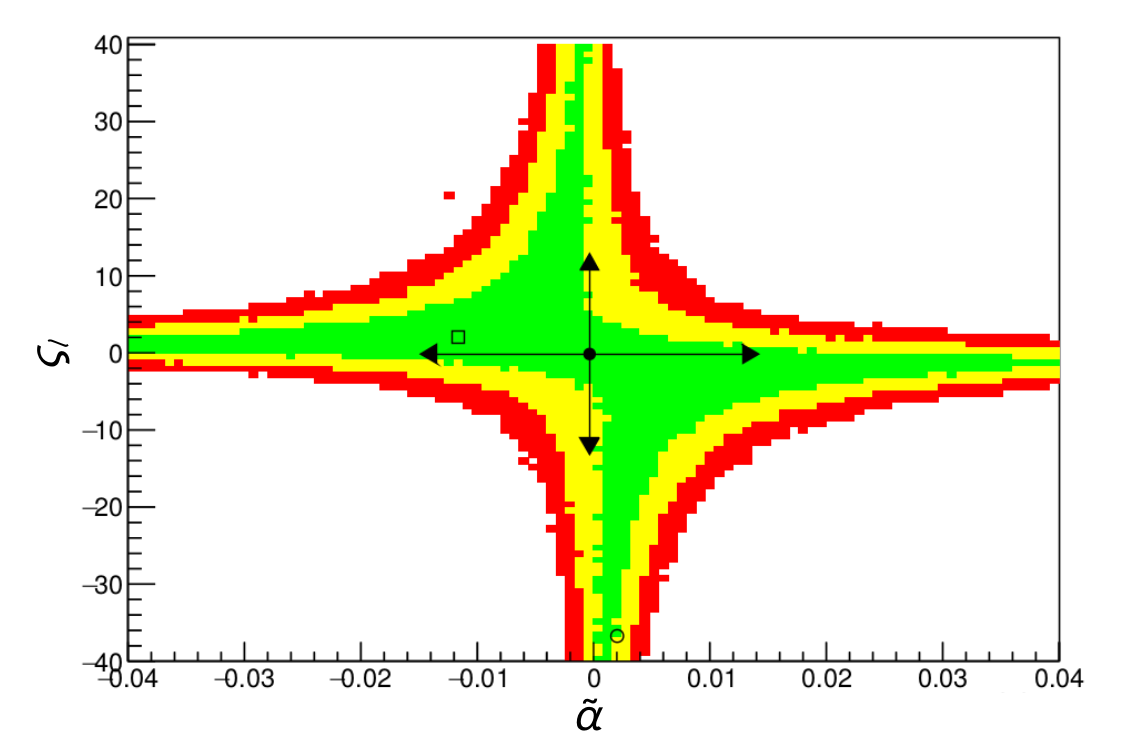}\\
\hline
Light $A$ & Light $H^\pm$ & Light $A$ \& $H^\pm$\\
\hline
    \end{tabular}
 }   
    \caption{Correlations between the mixing angle $\tilde\alpha$ and the alignment parameters $\varsigma_f$ in the three cases: a) light pseudoscalar, b) light charged scalar, c) pseudoscalar and charged scalar both lighter than the SM Higgs. The red, yellow and green regions represent $1\sigma$, $2\sigma$ and $3\sigma$ regions, respectively (LEP results are not included).}
    \label{fig:sig}
\end{figure}

We have used the open-source package {\tt HEPfit} \cite{DeBlas:2019ehy} to perform a global fit of the A2HDM. This package works within a Bayesian statistics framework. In addition to all the SM parameters, the A2HDM deals with ten extra inputs in the CP-conserving scenario: three alignment parameters in the Yukawa sector and seven parameters in the scalar potential. The scenario with all the new scalars being heavier than SM Higgs was already analysed in Refs. \cite{Karan:2023kyj, Eberhardt:2020dat, Karan:2023xze}. Here, we have studied the possibility of having light scalars. To be more specific, we have investigated three cases: a) light pseudoscalar, b) light charged scalar, c) pseudoscalar and charged scalar both lighter than the SM Higgs. The priors, necessary in Bayesian analyses, are chosen in suitable ranges. We depict the resulting correlations among the masses of the additional scalars for these three cases in Figure \ref{fig:mass}, while the correlations between mixing angle and alignment parameters are presented in Figure \ref{fig:sig}.

It is evident from both figures
that there exist viable parameter spaces for all the three light scenarios. However, there are a few observations that can be made. Firstly, BSM scalars with masses below 62.5 GeV seem to be ruled out. Though they become feasible in the light A \& $H^\pm$ (third) case, this occurs at the $3\sigma$ level only, and in isolated regions. What is happening is that masses smaller than $M_h/2$ open up the channels for the SM Higgs decaying into pairs of these BSM scalars, and this event changes the width (or invisible width) of the Higgs, which is highly constraining. In the light $A$ scenario the masses of the other two scalars remain very correlated and can go up to 360 GeV (at 3$\sigma$). In the light $H^\pm$ scenario, the masses of the other two scalars show a peculiar correlation indicating that if one becomes heavier, the other tends to become lighter, with two clearly preferred regions at $1\sigma$. In the third scenario, the light scalars show some correlation at $1\sigma$ and $2\sigma$ levels, and the heavy scalar can have masses till 600 GeV. Regarding the allowed values for the alignment parameters, the light $H^\pm$ scenario appears more restrictive for $\varsigma_{u}$ and $\varsigma_{d}$, and shows a smaller preference for $\varsigma_{\ell}-\tilde \alpha$ combinations with both simultaneously consistent with zero.

\section*{Acknowledgements}
This work has been supported by the Generalitat Valenciana (grant PROMETEO/2021/071), by MCIN/AEI/10.13039/501100011033 (grant No. PID2020-114473GB-I00), by the European Research Council (ERC) under the European Union’s Horizon 2020 research and innovation programme (grant agreement No. 949451),
and by a Royal Society University Research Fellowship (grant URF/R1/201553).


\begin{thebibliography}{99}

\bibitem{Branco:2011iw}
G.C.~Branco, P.M.~Ferreira, L.~Lavoura, M.N.~Rebelo, M.~Sher and J.P.~Silva,
\emph{{Theory and phenomenology of two-Higgs-doublet models}},
\href{https://doi.org/10.1016/j.physrep.2012.02.002}{\emph{Phys. Rept.} {\bfseries 516} (2012) 1} [\href{https://arxiv.org/abs/1106.0034}{{\ttfamily 1106.0034}}].
\vspace{-2mm}
\bibitem{Pich:2009sp}
A.~Pich and P.~Tuzon, \emph{{Yukawa Alignment in the Two-Higgs-Doublet Model}},
\href{https://doi.org/10.1103/PhysRevD.80.091702}{\emph{Phys. Rev. D} {\bfseries 80} (2009) 091702} [\href{https://arxiv.org/abs/0908.1554}{{\ttfamily 0908.1554}}].
\vspace{-2mm}
\bibitem{Braeuninger:2010td}
C.~B.~Braeuninger, A.~Ibarra and C.~Simonetto,
\emph{{Radiatively induced flavour violation in the general two-Higgs doublet model with Yukawa alignment}}, \href{https://doi.org/10.1016/j.physletb.2010.07.039}{\emph{Phys. Lett. B} \textbf{692} (2010) 189-195} [\href{https://arxiv.org/abs/1005.5706}{{\ttfamily 1005.5706}}].
\vspace{-2mm}
\bibitem{Jung:2010ik}
M.~Jung, A.~Pich and P.~Tuzon,
\emph{{Charged-Higgs phenomenology in the Aligned two-Higgs-doublet model}}, \href{https://doi.org/10.1007/JHEP11(2010)003}{\emph{JHEP} \textbf{11} (2010) 003} [\href{https://arxiv.org/abs/1006.0470}{{\ttfamily 1006.0470}}].
\vspace{-2mm}
\bibitem{Penuelas:2017ikk}
A.~Pe\~nuelas and A.~Pich,
\emph{{Flavour alignment in multi-Higgs-doublet models}}, \href{https://doi.org/10.1007/JHEP12(2017)084}{\emph{JHEP} \textbf{12} (2017) 084} [\href{https://arxiv.org/abs/1710.02040}{{\ttfamily 1710.02040}}].
\vspace{-2mm}
\bibitem{Ivanov:2015nea}
I.P.~Ivanov and J.P.~Silva, \emph{{Tree-level metastability bounds for the most general two Higgs doublet model}}, \href{https://doi.org/10.1103/PhysRevD.92.055017}{\emph{Phys. Rev. D} {\bfseries 92} (2015) 055017} [\href{https://arxiv.org/abs/1507.05100}{{\ttfamily 1507.05100}}].
\vspace{-2mm}
\bibitem{Ginzburg:2005dt}
I.~F.~Ginzburg and I.~P.~Ivanov,
\emph{{Tree-level unitarity constraints in the most general 2HDM}},
\href{https://doi.org/10.1103/PhysRevD.72.115010}{\emph{Phys. Rev. D} \textbf{72} (2005) 115010}
[\href{https://arxiv.org/abs/hep-ph/0508020}{{\ttfamily hep-ph/0508020}}].
\vspace{-2mm}
\bibitem{Celis:2013rcs}
A.~Celis, V.~Ilisie and A.~Pich,
\emph{{LHC constraints on two-Higgs doublet models}}, \href{https://doi.org/10.1007/JHEP07(2013)053}{\emph{JHEP} \textbf{07} (2013) 053 } [\href{https://arxiv.org/abs/1302.4022}{{\ttfamily 1302.4022}}].
\vspace{-2mm}
\bibitem{ParticleDataGroup:2024cfk}
S.~Navas \textit{et al.} [Particle Data Group],
\emph{{Review of particle physics}}, \href{https://doi.org/10.1103/PhysRevD.110.030001}{\emph{Phys. Rev. D} \textbf{110}, no.3, 030001 (2024)}.
\vspace{-2mm}
\bibitem{Celis:2013ixa}
A.~Celis, V.~Ilisie and A.~Pich,
\emph{{Towards a general analysis of LHC data within two-Higgs-doublet models}}, \href{https://doi.org/10.1007/JHEP12(2013)095}{\emph{JHEP} \textbf{12} (2013) 095} [\href{https://arxiv.org/abs/1310.7941}{{\ttfamily 1310.7941}}].
\vspace{-2mm}
\bibitem{ALEPH:2006tnd}
S.~Schael \textit{et al.} [ALEPH, DELPHI, L3, OPAL and LEP],
\emph{{Search for neutral MSSM Higgs bosons at LEP}}, \href{https://doi.org/10.1140/epjc/s2006-02569-7}{\emph{Eur. Phys. J. C} \textbf{47} (2006) 547-587} [\href{https://arxiv.org/abs/hep-ex/0602042}{{\ttfamily hep-ex/0602042}}].
\vspace{-2mm}
\bibitem{ALEPH:2013htx}
G.~Abbiendi \textit{et al.} [ALEPH, DELPHI, L3, OPAL and LEP],
\emph{{Search for Charged Higgs bosons: Combined Results Using LEP Data}},
\href{https://doi.org/10.1140/epjc/s10052-013-2463-1}{\emph{Eur. Phys. J. C} \textbf{73} (2013) 2463} [\href{https://arxiv.org/abs/1301.6065}{{\ttfamily 1301.6065}}].
\vspace{-2mm}
\bibitem{CMS}
CMS searches: \href{https://cms-results.web.cern.ch/cms-results/public-results/publications/index.html}{cms-results.web.cern.ch/cms-results/public-results/publications}
\vspace{-2mm}
\bibitem{ATLAS}
ATLAS searches: \href{https://twiki.cern.ch/twiki/bin/view/AtlasPublic/Publications}{twiki.cern.ch/twiki/bin/view/AtlasPublic/Publications}
\vspace{-2mm}
\bibitem{DeBlas:2019ehy}
J.~De~Blas et~al., \emph{{$\texttt{HEPfit}$: a code for the combination of indirect and direct constraints on high energy physics models}}, \href{https://doi.org/10.1140/epjc/s10052-020-7904-z}{\emph{Eur. Phys. J. C} {\bfseries 80} (2020) 456} [\href{https://arxiv.org/abs/1910.14012}{{\ttfamily 1910.14012}}].
\vspace{-2mm}
\bibitem{Karan:2023kyj}
A.~Karan, V.~Miralles and A.~Pich,
\emph{{Updated global fit of the ATHDM with heavy scalars}},
[\href{https://arxiv.org/abs/2307.15419}{{\ttfamily 2307.15419}}].
\vspace{-2mm}
\bibitem{Eberhardt:2020dat}
O.~Eberhardt, A.~Pe\~nuelas
and A.~Pich,
\emph{{Global fits in the Aligned Two-Higgs-Doublet model}},
\href{https://doi.org/10.1007/JHEP05(2021)005}{\emph{JHEP} {\bfseries 05} (2021) 005}
[\href{https://arxiv.org/abs/2012.09200}{{\ttfamily 2012.09200}}].
\vspace{-2mm}
\bibitem{Karan:2023xze}
A.~Karan, V.~Miralles and A.~Pich,
\emph{{Aligned two Higgs doublet model and the global fits}}, \href{https://doi.org/10.22323/1.449.0053}{\emph{PoS} \textbf{EPS-HEP2023} (2024) 053}.



\end{thebibliography}
\end{document}